\documentclass[runningheads]{llncs}
\pdfoutput=1
\usepackage{graphicx}
\usepackage{amsmath,amsfonts,graphicx,amssymb,subfig}
\usepackage{xcolor}
\usepackage{hyperref}
\makeatletter
\newcommand{\Vast}{\bBigg@{4}}
\makeatother
\definecolor{redblac}{rgb}{1, 0, 0}
\usepackage{changes}
\begin{document}
%
\title{Q-space quantitative diffusion MRI measures using a stretched-exponential representation} 
\titlerunning{Q-space quantitative diffusion MRI measures}
%

\authorrunning{T. Pieciak et al.}

\author{Tomasz Pieciak\inst{1,2}\and
Maryam Afzali\inst{3}\and
Fabian Bogusz\inst{2}\and 
Aja-Fern\'{a}ndez\inst{1}\and 
Derek K. Jones\inst{3}}

\institute{LPI, ETSI Telecomunicaci\'{o}n, Universidad de Valladolid, Valladolid, Spain\and
AGH University of Science and Technology, Krak\'{o}w, Poland \and
Cardiff University Brain Research Imaging Centre (CUBRIC), \mbox{School of Psychology, Cardiff University, Cardiff, United Kingdom}}
%

%
\maketitle              
\begin{abstract}
Diffusion magnetic resonance imaging (dMRI) is a relatively modern technique used to study tissue microstructure in a non-invasive way. Non-Gaussian diffusion representation is related to the restricted diffusion and can provide information about the underlying tissue properties. In this paper, we analytically derive $n$-th order statistics of the signal considering a stretched-exponential representation of the diffusion. Then, we retrieve the Q-space quantitative measures such as the Return-To-the-Origin Probability\index{Return-To-the-Origin Probability} (RTOP), Q-space mean square displacement\index{Q-space mean square displacement} (QMSD), Q-space mean fourth-order displacement\index{Q-space mean fourth-order displacement} (QMFD). The stretched-exponential representation enables the handling of the diffusion contributions from a higher $b$-value regime under a non-Gaussian assumption, which can be useful in diagnosing or prognosis of neurodegenerative diseases in the early stages. 
Numerical implementation of the method is freely available at \url{https://github.com/TPieciak/Stretched}.
\end{abstract}

\section{Introduction}  
Magnetic resonance imaging (MRI) is a powerful technique in clinical applications for diagnoses or prognoses of several diseases in the central nervous system \cite{basser1996microstructural,kubicki2007review}. 
Diffusion-weighted MRI (dMRI) modality is sensitive to the random motion of water molecules in the tissue, and it is vastly used in both clinical and basic science to characterize tissue water behavior. 
In the category of dMRI studies, different imaging techniques can extract the microstructural features of the tissue, such as size, shape, and anisotropy. Many studies have shown a relationship between the changes in the diffusion properties of the tissue and the relevant alteration in the underlying tissue microstructure \cite{rovaris2007diffusion}. 

One of the most important features of dMRI is its sensitivity to the anisotropy in the tissue. Diffusion tensor imaging (DTI) \cite{basser1996microstructural} is the most common technique in clinical studies. The spins displacement in DTI is assumed to be Gaussian distributed, and some scalar anisotropy indices such as fractional anisotropy (FA), axial, and radial diffusivity (AD, RD), and mean diffusivity (MD) were defined directly from second-order tensor representation \cite{basser1995inferring,westin2002processing}. The most common and simplest assumption in dMRI is the Gaussian assumption \cite{jones2013white} on the spin displacement, which results in a mono-exponential decay of a diffusion signal versus the $b$-value parameter \cite{callaghan2011translational}. The Gaussian assumption in DTI is valid when the media is a simple fluid, and we have free diffusion. In more complicated structures such as tissue, there are restrictions in water diffusion, and therefore, the decay is much slower than a mono-exponential decay \cite{bennett2003characterization}.

The presence of non-mono-exponential decay shows that the diffusion is restricted by the underlying microstructure of the tissue. If the structure's size is similar to the diffusion length-scale, then the diffusion deviates from the Gaussian towards a non-Gaussian displacement regime \cite{grebenkov2008laplacian}. Therefore diffusion MRI can be used to probe the microstructural properties of the underlying tissue geometry by methods such as bi-exponential \cite{clark2000water}, stretched-exponential \cite{bennett2003characterization}, composite hindered and restricted model of diffusion (CHARMED) \cite{assaf2005composite}, AxCaliber \cite{assaf2008axcaliber}, ActiveAx \cite{alexander2008general} or neurite orientation dispersion and density imaging (NODDI) \cite{zhang2012noddi}. All these methods can be used to investigate tissue geometry, but they are not all equally applicable in all situations. 
Other methods such as the high angular resolution diffusion imaging (HARDI) \cite{ozarslan2006resolution,tuch2003diffusion} and diffusion kurtosis imaging (DKI) \cite{jensen2005diffusional} were also proposed. To obtain the non-Gaussian property of the signal higher $b$-values are required \cite{le1991molecular}. 

Alternative to model the tissue's underlying properties is the ensemble average propagator (EAP), which represents the probability that the water molecule moves in a specific direction under a certain diffusion time \cite{descoteaux2011multiple,ozarslan2013mean}. From the EAP representation, one can retrieve different Q-space quantitative measures such as the Return-To-the-Origin Probability (RTOP), Q-space mean square displacement (QMSD), or Q-space mean fourth-order displacement (QMFD). For instance, the RTOP measure is shown to be a useful index for cellularity and diffusion restrictions \cite{avram2016clinical}, while QMSD and QMFD are sensitive to contributions from slow or restricted diffusion \cite{ning2015estimating}. 

Different methods have been proposed so far to estimate the EAP and EAP-related features such as the multiple q-shell diffusion propagator Imaging (mq-DPI) \cite{descoteaux2011multiple}, Hybrid Diffusion Imaging (HYDI) \cite{wu2008computation}, Mean Apparent Propagator \index{Mean Apparent Propagator} (MAP-MRI) \cite{ozarslan2013mean}, Radial Basis Functions (RBFs) \cite{ning2015estimating} Laplacian-regularized MAP-MRI (MAPL) \cite{fick2016mapl}, and Generalized Diffusion Spectrum MRI (GDSI) \cite{tian2019generalized}. These techniques are typically computationally intensive or require a huge amount of densely sampled Cartesian or multiple-shell data to estimate the EAP and its related features correctly. Recently, a single-shell technique that can estimate micro-structure diffusion scalar measures directly from the data has been proposed \cite{aja2020micro,pieciak2019single}. This approach, although it enables one to estimate the measures rapidly and directly from the data, assumes a Gaussian profile of the signal, thus it might be problematic to recover higher $b$-value contributions to the signal. 

In this paper, we analytically derive the $n$-th order statistics of the signal considering a stretched-exponential decay to represent the Gaussian and non-Gaussian parts of the signal. In practice, when no information about the number of compartments is provided, the stretched exponential is a good choice  \cite{bennett2003characterization}. Given the general formulation in the Q-space domain, we obtain closed-form formulas to retrieve basic indexes such as the RTOP, QMSD, or QMFD directly from the data in a manner analogous to direct techniques \cite{aja2020micro,pieciak2019single,wu2008computation}. However, the proposal is no longer limited by a Gaussian assumption and can be used to retrieve the diffusion contributions under the higher $b$-values regime.

\section{Theory}
In this section, we start with the definition of the EAP and diffusion MR signal representation using a stretched-exponential function, and then we use this representation to extract the Q-space scalar measures such as the RTOP, QMSD, and QMFD.

\subsection{Diffusion MR signal representation}
The ensemble average propagator\index{ensemble average propagator} (EAP) is a three-dimensional probability density function that represents the average displacement of spins during the diffusion time. The EAP, $P(\mathbf{R})$, is related to the diffusion MR signal attenuation $E(\mathbf{q})$ via the Fourier transform \cite{descoteaux2011multiple,ozarslan2013mean,wu2008computation}
\begin{equation}
 P(\mathbf{R}) = \int_{\mathbb{R}^3} E(\mathbf{q}) \exp(-j 2\pi  \mathbf{q}^T \mathbf{R}) d^3\mathbf{q}, \ \ \ j^2 = -1,
 \label{eq:EAP_function}
\end{equation}
with $E(\mathbf{q}) = S(\mathbf{q})/S(0)$ being the normalized diffusion signal, $S(\mathbf{q})$ is the diffusion signal acquired at wave vector $\mathbf{q}$, $S(0)$ is the baseline measured without a~diffusion sensitization. 

The signal in Eq. (\ref{eq:EAP_function}) can be represented by a mono-exponential decay $E(\mathbf{g}) = \exp\left(  -b \mathbf{g}^T \mathbf{D} \mathbf{g} \right)$ with $\mathbf{g}$ being a normalized vector $\mathbf{g} = \mathbf{q}/\| \mathbf{q} \|$ and $\mathbf{D}$ is a covariance matrix of a Gaussian EAP or a more general Kohlrausch--Williams--Watts function so-called a stretched-exponential representation given by \cite{bennett2003characterization,magin2019fractional,williams1970non}
\begin{equation}
E(\mathbf{g}) = \exp\left(  -(b D(\mathbf{g}))^{\alpha(\mathbf{g})} \right), \ \ \ \alpha(\mathbf{g}) \in (0, 1]
\label{eq:stretched}
\end{equation}
with the so-called the $b$-value $b=4\pi^2 \tau \| \mathbf{q} \|^2$ $[\mathrm{s}/\mathrm{mm}^2]$ with $\tau = \Delta - \delta/3$ $[\mathrm{s}]$ being the effective diffusion time, $D(\mathbf{g})$ and $\alpha(\mathbf{g})$ being the apparent diffusion and stretching parameters at direction $\mathbf{g}$, respectively. Notice here once the stretching parameter tends to unitary, i.e., \mbox{$\alpha(\mathbf{g}) \rightarrow 1$}, the stretched-exponential representation (\ref{eq:stretched}) reduces to a mono-exponential signal\index{mono-exponential signal} decay.

\subsection{Q-space domain quantitative measures}
\label{subsec:measures}
In what follows, we analytically derive $n$-th order statistics of the stretched-exponential representation given by the Eq. (\ref{eq:stretched}). This enables to easily retrieve three quantitative Q-space measures namely the RTOP [$\mathrm{mm}^{-3}$] being the probability in the origin, $P(\mathbf{0})$, QMSD [$\mathrm{mm}^{-5}$] and QMFD [$\mathrm{mm}^{-7}$] defined as the second- and fourth-order statistics of the signal $E(\mathbf{q})$ respectively \cite{descoteaux2011multiple,ning2015estimating} 
\begin{equation}
 \mathrm{RTOP}\hspace{-0.2mm} = \hspace{-0.2mm}\int_{\mathbb{R}^3}\hspace{-0.4mm} E(\mathbf{q}) d^3\mathbf{q}, \ \ \mathrm{QMSD}\hspace{-0.2mm} = \hspace{-0.2mm}\int_{\mathbb{R}^3} \| \hspace{-0.4mm} \mathbf{q} \|^2 E(\mathbf{q}) d^3\mathbf{q}, \  \ \mathrm{QMFD}\hspace{-0.2mm} = \hspace{-0.2mm}\int_{\mathbb{R}^3} \hspace{-0.4mm} \| \mathbf{q} \|^4 E(\mathbf{q}) d^3\mathbf{q},
 \label{eq:rtop_qmsd_qmfd}
\end{equation}
where $\| \cdot  \|$ is the vector norm of the wave vector $\mathbf{q}$.

We specify now a more general equation in the Q-space domain related to the $n$-th order statistics of the signal attenuation $E(\mathbf{q})$. Considering the stretched-exponential representation of the signal (\ref{eq:stretched}) and a spherical coordinate system $(q, \theta, \varphi)$ with polar $\theta$ and azimuthal $\varphi$ angles, and a radial coordinate $q = \| \mathbf{q} \|$ $[\mathrm{mm}^{-1}]$ we define the $n$-th order statistics of the signal attenuation $E(\mathbf{q})$
\begin{equation}
\begin{split}
M_n &= \int_{\mathbb{R}^3} \| \mathbf{q} \|^n \exp\left(  -(4\pi^2 \tau \| \mathbf{q} \|^2 D(\mathbf{g}))^{\alpha(\mathbf{g})} \right) d^3\mathbf{q}\\
    &= \int_{0}^{2\pi}  \int_{0}^{\pi}  \int_{0}^{\infty} \exp\left(-\left( 4\pi^2\tau q^2 D(\theta, \varphi)\right)^{\alpha(\theta, \varphi)}\right) q^{n+2} \sin \theta\, dq\, d\theta\, d\varphi,
\end{split}
\label{eq:Mn}
\end{equation}\\
where $D(\theta, \varphi)$ and $\alpha( \theta, \varphi)$ are the apparent diffusion coefficient and stretching parameter both defined in the spherical coordinate system. Next, we rewrite the integral (\ref{eq:Mn}) as follows (see Gradshteyn \& Ryzhik \cite{gradshteyn2014table}, p. 370, Eq. 3.478(1))
\begin{equation}
\begin{split}
  M_n &= C^\tau_n \int_{0}^{2\pi} \int_{0}^{\pi} \mathrm{\Gamma}\left(\frac{n+3}{2\alpha(\theta, \varphi)}\right)\alpha^{-1}(\theta, \varphi) \, D^{-(n+3)/2}(\theta, \varphi) \sin\theta\, d\theta\, d\varphi\\
      &= C^\tau_n  \iint_{\Sigma} \mathrm{\Gamma}\left(\frac{n+3}{2\alpha(\theta, \varphi)}\right)\alpha^{-1}(\theta, \varphi) \, D^{-(n+3)/2}(\theta, \varphi) \, d\Sigma,
\end{split} 
 \label{eq:Mn_spherical}
\end{equation}
where $C^\tau_n = 2^{-n-4} \pi^{-n-3} \tau^{-(n+3)/2} $ is a diffusion time dependent constant and $\mathrm{\Gamma}(\cdot)$ is the gamma function. Notice here that the last equation is a surface integral over the surface with a unitary radius, i.e., $q=1$.

\subsection{Numerical implementation}
To evaluate the surface integral (\ref{eq:Mn_spherical}), one can assume the surface area element, $\Delta \Sigma$, is inversely proportional to the number of sampled data points 
(e.g., the number of evenly distributed directions $N_g$, $\Delta \Sigma  =  4\pi/N_g$). Transforming the Eq. (\ref{eq:stretched}) the diffusion becomes $D(\mathbf{g}) = 4^{-1}\pi^{-2} \tau^{-1} q^{-2}\left( - \log E(\mathbf{q})\right)^{1/\alpha(\mathbf{g})}$, and~thus the Eq.~(\ref{eq:Mn_spherical}) can be rewritten in the following form
\begin{equation}
 M_n^{(1)} = \frac{1}{2} q^{n+3} \left< \mathrm{\Gamma}\left(  \frac{n+3}{2\alpha(\mathbf{g})} \right)  \alpha^{-1} (\mathbf{g})\left( - \log E(\mathbf{q})\right)^{- \frac{n+3}{2\alpha(\mathbf{g})}} \right>_{\mathbf{q} \in \mathbb{S}^2}
 \label{eq:direct}
\end{equation}\\
with 
$\left< \cdot \right>_{\mathbf{q} \in \mathbb{S}^2}$ being a direction-averaged signal over a single acquisition shell. 
Notice that the Eq. (\ref{eq:direct}) can be evaluated using the samples retrieved from the resampled data to uniformly cover the surface (e.g., the spherical harmonics \cite{aja2020micro}).

The numerical reciprocal of the negative log-diffusion function given in Eq.~(\ref{eq:direct}) might be prone to instabilities, i.e., the signal attenuation $E(\mathbf{q}) \to 1$ the function $(-\log(E(\mathbf{q})))^{-1} \to \infty$ \cite{aja2018scalar,pieciak2019single}. 
Alternatively, one can refine (\ref{eq:direct}) to incorporate its second-order series expansion. To this end, we define a twice differentiable function $f\colon \mathbb{R} \rightarrow \mathbb{R}$ given by $f(X) = \mathrm{\Gamma}\left(  \frac{n+3}{2\alpha} \right)  \alpha^{-1} X ^{-(n+3)/(2\alpha)}$ with $n\geq 0$. The second-order series expansion of the expectation of the function $f(X)$ around the expectation $\mathbb{E} \left\{ X \right\}$ is given then by $\mathbb{E} \left\{ f(X) \right\} \approx f\left(\mathbb{E} \left\{ X \right\}\right) + \left.\frac{1}{2} \frac{d^2 f}{d X^2}\right|_{X=\mathbb{E} \left\{ X \right\}} \cdot \mathrm{Var}\left\{ X \right\}$. After using some algebra we arrive at the following closed-form formula
\begin{equation*}
 \begin{split}
\mathbb{E} \left\{ f(X) \right\} &\approx \frac{1}{8}\mathrm{\Gamma}\left(  \frac{n+3}{2\alpha} \right)\alpha^{-3}\, \mathbb{E} \left\{ X \right\}^{-\frac{n+3}{2\alpha}} \Big( (n+3)(n+3+2\alpha)\mathbb{E} \left\{ X^2 \right\}\mathbb{E} \left\{ X \right\}^{-2} \\ 
& \hspace{4.2cm} + 8\alpha^2 - (n+3)(n+3+2\alpha)\Big).
 \end{split}
 \label{eq:second_order_expansion} 
\end{equation*}\\
Given a stretched-exponential decay \index{stretched-exponential decay} again (\ref{eq:stretched}) and a second-order series expansion of the expectation, we define an approximation to the measure  (\ref{eq:direct}) 
\begin{equation}
\begin{split}
 M_n^{(2)} &=   \frac{1}{2} q^{n+3} \left< \frac{1}{8} \mathrm{\Gamma}\left(  \frac{n+3}{2\alpha(\mathbf{g})} \right)  \alpha^{-3} (\mathbf{g}) \right>_{\mathbf{q} \in \mathbb{S}^2} \Big \langle - \log E(\mathbf{q}) \Big \rangle^{-\left< \frac{n+3}{2\alpha(\mathbf{q})} \right>_{\mathbf{q} \in \mathbb{S}^2}}_{\mathbf{q} \in \mathbb{S}^2}  \\
     & \times \Vast[\frac{(n + 3) \Big \langle (n + 3)\mathbf{1} +  2\alpha(\mathbf{g})  \Big \rangle_{\mathbf{q} \in \mathbb{S}^2}   \left<  \left( - \log E(\mathbf{q})\right)^2  \right>_{\mathbf{q} \in \mathbb{S}^2}}{\Big \langle   - \log E(\mathbf{q})  \Big \rangle_{\mathbf{q} \in \mathbb{S}^2}^2} \\
     & \hspace{2.5cm} + \Big \langle  8\alpha^2(\mathbf{g}) - 2(n+3)\alpha(\mathbf{g}) - (n+3)^2\mathbf{1} \Big \rangle _{\mathbf{q} \in \mathbb{S}^2} \Vast]
\end{split} 
 \label{eq:final}
\end{equation}\\
with $\mathbf{1}$ being the all-ones vector. Here, we have simplified our derivations in the series expansion procedure; thus, we direction average the stretching parameter to obtain the final formula. From Eq. (\ref{eq:final}) we can define basic Q-space domain measures such as the RTOP ($M_0$), QMSD ($M_2$) or QMFD  ($M_4$; see Eq. (\ref{eq:rtop_qmsd_qmfd})). 
The proposed stretched-exponential method requires a multiple-shell acquisition with at least two-shells at different $b$-values to fit the representation given by Eq. (\ref{eq:stretched}). Once the representation is fitted, a single-shell data at a fixed $b$-value is used to calculate the measures.
In section 2.4, we define a simple optimization cost function to retrieve the stretched representation of the diffusion signal.

\subsection{Optimization of stretched-exponential representation}
 To retrieve a stretched-exponential representation at direction $\mathbf{g}$ of the diffusion, we define an optimization cost function and solve it using a non-linear least squares \index{non-linear least squares} procedure
\begin{equation*}
    \left( D(\mathbf{g}), \alpha(\mathbf{g}) \right) = \underset{D'(\mathbf{g}), \, \alpha'(\mathbf{g})}{\mathrm{argmin}}  \ \frac{1}{2}\sum_{\mathbf{q}\colon \mathbf{q} \parallel \mathbf{g}} \Big[ S(\mathbf{q})  - S(0) \exp\left(  -(4\pi^2 \tau \| \mathbf{q} \|^2 D'(\mathbf{g}))^{\alpha'(\mathbf{g})} \right) \Big]^2.
    \label{eq:numerical_optimization}
\end{equation*}\\
We used a bound-constrained minimization via the trust-region reflective method with a linear loss function to find the optimal parameters. Notice here the procedure (\ref{eq:numerical_optimization}) applies for each direction $\mathbf{g}$ independently and might use the only a~subset of $\mathbf{q}$-values employed to acquire the data.  

\section{Materials and methods}
In this study, we used \emph{ex vivo} rat brain data as well as \emph{in vivo} human brain data that was publicly available by Hansen et al. \cite{hansen2016data}.
\subsection{\emph{Ex vivo} rat brain data}
The \emph{ex vivo} data were collected using a Bruker Biospec 9.4T (Bruker Biospin, Germany) with a 15 mm quadrature coil. Diffusion-weighted images were acquired in 15 $b$-value shells ranging from 0 to 5000 $\mathrm{s}/\mathrm{mm}^2$ with a step size of 200 $\mathrm{s}/\mathrm{mm}^2$ and 33 directions per each shell utilizing a spin echo sequence. Fifteen axial slices were collected at a resolution of $100 \times 100 \times 500 \ \mu \mathrm{m}^3$, matrix size $128 \times 128$, echo time of $\mathrm{TE} = 23.3$ ms, repetition time of $\mathrm{TR} = 4$ s, and diffusion timing of $\delta/\Delta = 4/14$ ms. The data set was averaged twice to improve the signal-to-noise ratio being around 75 at the baseline. 
\subsection{\emph{In vivo} Human brain data}
One healthy participant was scanned in an \emph{in vivo} study using a Siemens Trio 3T equipped with a 32 channel head coil. The protocol comprised one $b = 0$ and 15 non-zero shells ranging from 200 $\mathrm{s}/\mathrm{mm}^2$ to 3000 $\mathrm{s}/\mathrm{mm}^2$ with the step size of 200 $\mathrm{s}/\mathrm{mm}^2$ and 33 directions per shell. Nineteen axial slices with a voxel size of $2.5$ mm isotropic and a $96 \times 96$ matrix size, $\mathrm{TE} = 116$ ms, $\mathrm{TR} = 7200$ ms, $\mathrm{TI} = 2100$ ms were obtained. The diffusion timings were estimated to be $\delta/\Delta = 29/58$ ms. The SNR of the baseline signal is around 39. In our experiments we used a five-shell acquisition with $200$, $1000$, $1800$ $2400$, and $3000 \ \mathrm{s}/\mathrm{mm}^2$.

\subsection{Comparison to the Q-space measures from different methods}
In subsection \ref{subsec:measures} we introduced three measures that is the RTOP, $\mathrm{QMSD}$ and $\mathrm{QMFD}$. In this work, we evaluate the proposed stretched-exponential Q-space measures and compare them to those obtained from the MAP-MRI technique \cite{ozarslan2013mean} (positivity constraint), MAPL \cite{fick2016mapl} (regularization parameter $\lambda=0.2$), RBF \cite{ning2015estimating} ($l_1$ regularization with $\lambda=0.00055$), 3D-SHORE \cite{ozarslan2009simple,zucchelli2016lies} (scale factor $\zeta=1/( 8\pi^2 \tau D)$), and a single-shell approach  \cite{aja2020micro,pieciak2019single}. Except for the aforementioned frameworks, we calculate also the RTOP measure directly from diffusion tensor eigenvalues (a non-linear least squares fitting via the Levenberg-Marquardt method) as $\mathrm{RTOP} = \left( 3\pi \tau\right)^{-3/2} \left( \lambda_1 \lambda_2 \lambda_3\right)^{-1/2}$.

\section{Results and discussion}
\begin{figure*}[!b]
  \centering
  \includegraphics[width=0.72\textwidth]{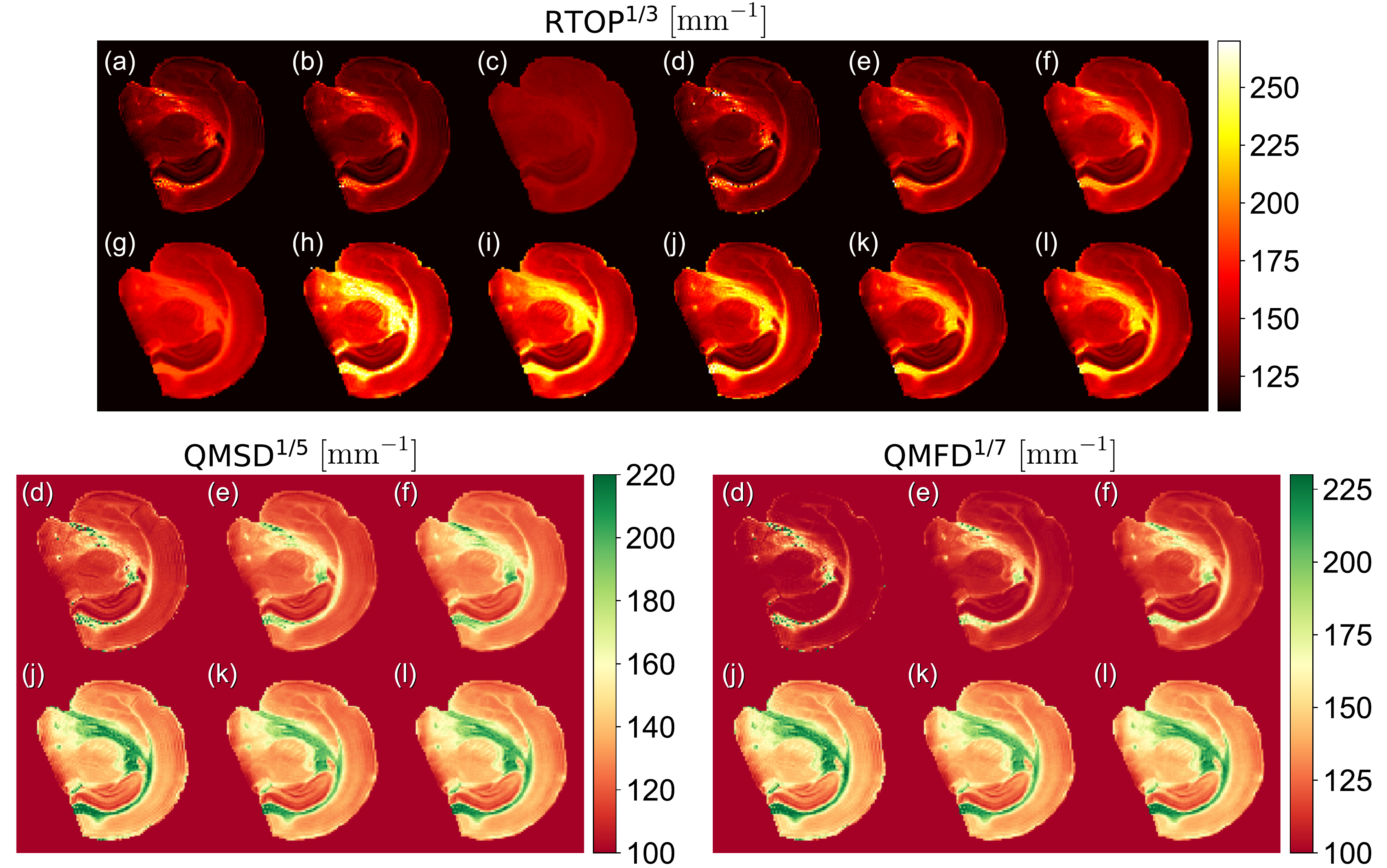}  
  \caption{Visual inspection of the RTOP, QMSD and QMFD measures on \emph{ex vivo} rat brain data: (a)  DTI ($b=1000 \ \mathrm{s}/\mathrm{mm}^2$), (b) DTI ($b=1400 \ \mathrm{s}/\mathrm{mm}^2$), (c) RBF, (d) single-shell ($b=1000 \ \mathrm{s}/\mathrm{mm}^2$), (e) single-shell ($b=3000 \ \mathrm{s}/\mathrm{mm}^2$), (f) single-shell ($b=5000 \ \mathrm{s}/\mathrm{mm}^2$), (g) 3D-SHORE, (h) MAP-MRI, (i) MAPL, (j) stretched-exponential ($b=1000 \ \mathrm{s}/\mathrm{mm}^2$), (k) stretched-exponential ($b=3000 \ \mathrm{s}/\mathrm{mm}^2$) and (l) stretched-exponential ($b=5000 \ \mathrm{s}/\mathrm{mm}^2$). }  
  \label{fig:visual_experiment} %
\end{figure*} 
%
\begin{figure*}[!t]
  \centering
  \includegraphics[width=0.8\textwidth]{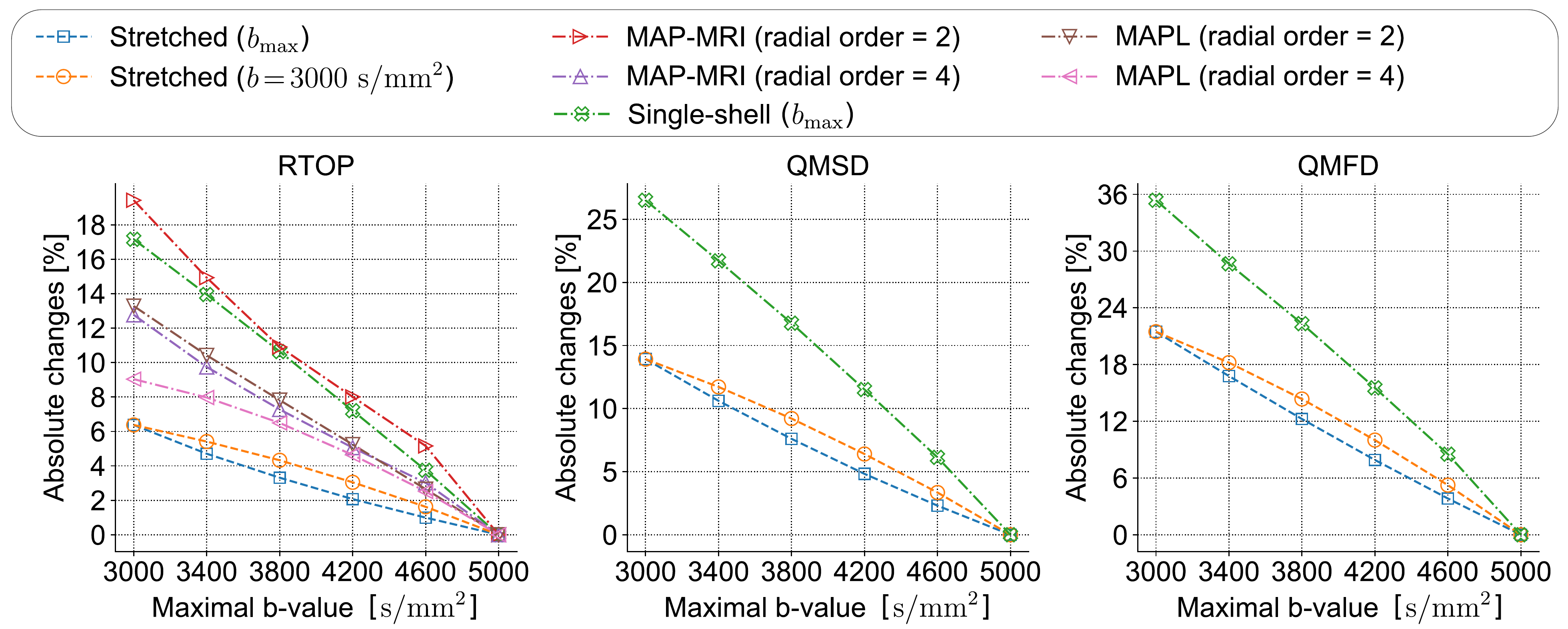}  
  \caption{The mean absolute changes of the RTOP, QMSD and QMFD measures in \emph{ex vivo} rat brain data in  terms of maximal $b$-value ($b_\mathrm{max}$) used to estimate the EAP/calculate the measure under different techniques. For stretched-exponential representation two variants are used in measure calculation process namely \mbox{$b$-value} at $b_\mathrm{max}$ and $b=3000 \ \mathrm{s}/\mathrm{mm}^2$.}  
  \label{fig:simulation_rat} %
\end{figure*} 
%
%
\begin{figure*}[!t]
  \centering
  \includegraphics[width=0.8\textwidth]{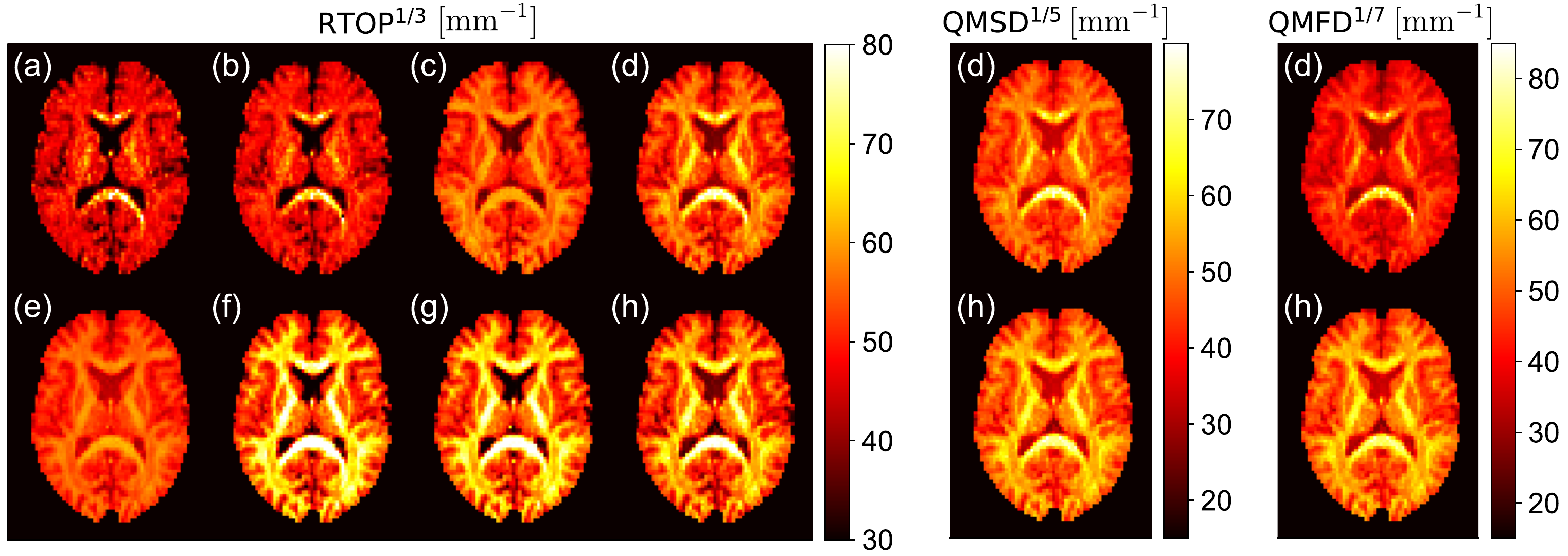}  
  \caption{Visual inspection of the measures on \emph{in vivo} human brain data estimated using various approaches: (a)  DTI ($b=1000 \ \mathrm{s}/\mathrm{mm}^2$), (b) DTI ($b=1400 \ \mathrm{s}/\mathrm{mm}^2$), (c) RBF, (d) single-shell ($b=3000 \ \mathrm{s}/\mathrm{mm}^2$), (e) 3D-SHORE, (f) MAP-MRI, (g)~MAPL and (h) stretched-exponential ($b=3000 \ \mathrm{s}/\mathrm{mm}^2$).} 
  \label{fig:visual_human3} %
\end{figure*} 
%
%
%
%
In the first experiment, we visually evaluate the measures using \emph{ex vivo} rat brain data retrieved using different methodologies namely the DTI (at $1000$ and $1400 \ \mathrm{s}/\mathrm{mm}^2$), RBF, 3D-SHORE, MAP-MRI, MAPL (the EAPs in all four are fitted to a three-shells acquisition, i.e., $b=1000$, $3000$ and $5000 \ \mathrm{s}/\mathrm{mm}^2$), single-shell technique, and proposed stretched-exponential one (see Fig. \ref{fig:visual_experiment}). For the proposed technique, we use three-shells for fitting the representation while a single-shell to calculate the measure. 
Visually inspecting the RTOP measure shows that DTI based ones (Fig. \ref{fig:visual_experiment}(a, b)) and single-shell technique (Fig. \ref{fig:visual_experiment}(d--f)) return smaller values compared to all other methods. The RBF and 3D-SHORE have the lowest contrast between the white matter and gray matter tissue in the measure (Fig. \ref{fig:visual_experiment}(c, g)). In the single-shell method by increasing the $b$-value from $1000$ to $5000 \  \mathrm{s}/\mathrm{mm}^2$, the RTOP value increases (Fig. \ref{fig:visual_experiment}(d--f, RTOP)). Comparing the EAP-based techniques and stretched-exponential representations (Fig. \ref{fig:visual_experiment} (c, g--l, RTOP)), 3D-SHORE and RBF provide the lowest while MAP-MRI gives the highest intensity in white matter areas. MAPL results of RTOP are similar to MAP-MRI while the RTOP values in MAPL are slightly lower than the MAP-MRI. Our proposed method of stretched-exponential provides similar RTOP maps for different $b$-values ($b = 1000$, $3000$, and $5000 \ \mathrm{s}/\mathrm{mm}^2$) and it preserves the consistency of the measures between different $b$-values which is not observed in single-shell technique (Fig. \ref{fig:visual_experiment}(d--f, j--l, RTOP)). Clearly, introducing the stretched-exponential representation enabled to improve contrast and kept the RTOP measure's uniformity across the $b$-values.
In the QMSD/QMFD measures, single-shell method at $b = 1000 \ \mathrm{s}/\mathrm{mm}^2$ has the lowest value in the both gray matter and white matter while the $b = 5000 \ \mathrm{s}/\mathrm{mm}^2$ has the highest and $b = 3000 \ \mathrm{s}/\mathrm{mm}^2$ is the intermediate between the three alternatives of the single-shell method (Fig. \ref{fig:visual_experiment} (d--f, QMSD/QMFD). Notice here that again the behavior of the QMSD/QMFD measures across the $b$-values are preserved, and the proposed stretched-exponential representation keeps the consistency of the quantities while changing the $b$-value used to calculate the measure.

\begin{figure*}[!t]
  \centering
  \includegraphics[width=0.9\textwidth]{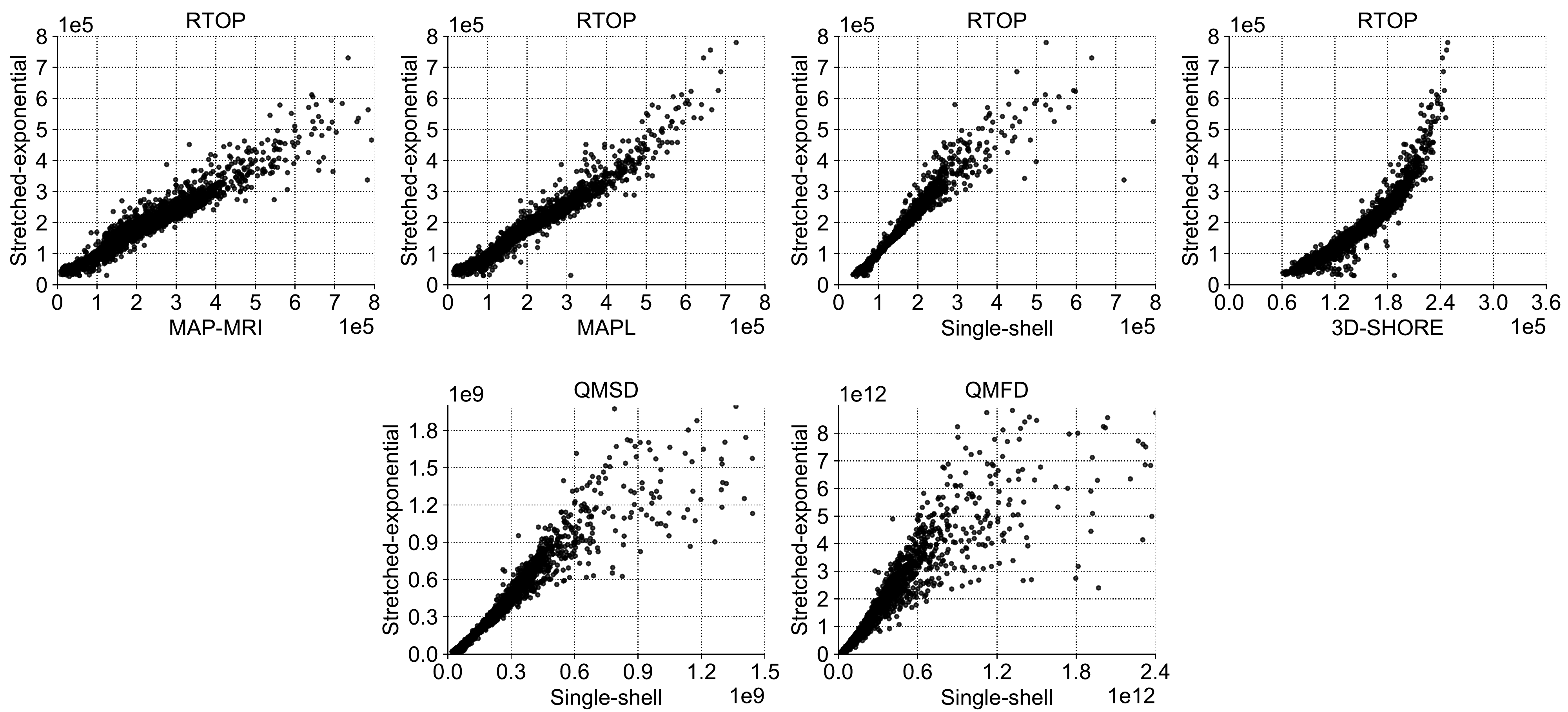}  
  \caption{Correlations between the RTOP retrieved from \emph{in vivo} human brain data using the proposed stretched-exponential and state-of-the-art methods namely MAP-MRI, MAPL, single-shell and 3D-SHORE (top), and correlations between the proposal and single-shell approach in case of QMSD/QMFD (bottom).}  
  \label{fig:correlations_measures} %
\end{figure*} 

In the second experiment, we extrapolate the previous one and evaluate the absolute changes in the measures due to the changes in the maximal $b$-value parameter. In this experiment we used six different acquisitions from three-shells (i.e., $b=1000$, $2000$ and $b=3000 \ \mathrm{s}/\mathrm{mm}^2$) to eight-shells (up to $b=5000 \ \mathrm{s}/\mathrm{mm}^2$ with a step of $400 \ \mathrm{s}/\mathrm{mm}^2$). For the proposed methodology we fit the representation using $k$ shells ($k=3,\ldots,8$; similarly to the EAP-based methods) while calculate the measures using only one shell, i.e., $b=3000 \ \mathrm{s}/\mathrm{mm}^2$ or $b_\mathrm{max}$. Fig.~\ref{fig:simulation_rat} depicts the mean absolute changes of the RTOP, QMSD, and QMFD measures in terms of maximal $b$-value ($b_\mathrm{max}$) under different methodologies used to estimate the measures. For the estimation of RTOP, our proposed stretched-exponential methodology has the minimum mean absolute changes for both $b_\mathrm{max}$ and $b = 3000 \ \mathrm{s}/\mathrm{mm}^2$ used to retrieve the measure once fitted the Eq. (\ref{eq:numerical_optimization}). It is worth noticing here that the single-shell technique is heavy load due to the changes in the maximal $b$-value and our proposal improved the results though the measure is still calculated from a single-shell. 
As for the two other measures, our proposed method is superior to the single-shell method, while again, the one with $b_\mathrm{max}$ is slightly better than that with $b = 3000 \ \mathrm{s}/\mathrm{mm}^2$.

\begin{table}[!t]
\renewcommand{\arraystretch}{1.2}
        \begin{center}
        \scriptsize
            \begin{tabular}{l|c|c|c|c|c|c|c|c}
                        \hline\hline
                         & DTI & MAP-MRI & MAPL & Single (1) & Single (2) & 3D-SHORE & SE (1) & SE (2) \\
                        \hline
                        DTI & $\times$ &  &  &  &  &  &  &  \\
                        \hline
                        MAP-MRI & 0.681 & $\times$ &  &  &  &  &  &  \\
                        \hline
                        MAPL & 0.674 & 0.981 & $\times$ &  &  &  &  &  \\
                        \hline
                        Single (1) & 0.827 & 0.901 & 0.895 & $\times$ &  &  &  &  \\
                        \hline
                        Single (2) & 0.797 & 0.936 & 0.932 & 0.978 & $\times$ &  &  &  \\
                        \hline
                        3D-SHORE & 0.642 & 0.928 & 0.961 & 0.863 & 0.895 & $\times$ &  &  \\
                        \hline
                        SE (1) & 0.739 & 0.951 & 0.965 & 0.944 & 0.954 & 0.931 & $\times$ &  \\
			            \hline
			            SE (2) & 0.728 & 0.957 & 0.973 & 0.925 & 0.959 & 0.933 & 0.982 & $\times$ \\
                        \hline\hline
                \end{tabular}
        \end{center}
        \caption{Pearson's correlation coefficient between different methodologies used to retrieve the RTOP measure from human brain data (see Fig. \ref{fig:visual_human3}). Legend: Single (1), (2) -- single-shell technique at $b = 2400$ and $b = 3000 \ \mathrm{s}/\mathrm{mm}^2$; SE (1), (2) -- stretched-exponential at $b = 2400$ and $b = 3000 \ \mathrm{s}/\mathrm{mm}^2$, respectively.}
        \label{tab:table1}
\end{table}

Fig.~\ref{fig:visual_human3} is devoted to visual inspection of the RTOP and QMSD/QMFD measures on \emph{in vivo} human brain data estimated using various state-of-the-art approaches. The observed trend in the \emph{in vivo} maps of RTOP is similar to the one observed in \emph{ex vivo} data (Fig. \ref{fig:visual_experiment}; RTOP). 
Again, a comparable behavior of the QMSD/QMFD measures can be observed, i.e., the single-shell technique generally exhibits smaller values of the measure in white matter areas than stretched-exponential representation. Notice here that the QMFD measure obtained from the single-shell technique cannot catch heavy tails of the signal distribution properly from the data. 

Lastly, we extrapolate the previous experiment using \emph{in vivo} human data and evaluate the correlation between measures both visually and numerically. Here, we again use a five-shells acquisition to calculate the measures. 
Fig.~\ref{fig:correlations_measures}(top row) illustrates correlations between the RTOP retrieved using the proposed approach and those obtained under different state-of-the-art methods, namely MAP-MRI, MAPL, the single-shell approach, and 3D-SHORE over the image mask (white matter, gray matter, and cerebrospinal fluid areas). Our method shows a good correlation with all the other four methods being characterized by Pearson's correlation coefficient equal to $\rho=0.925$ is the worst case (see Table~\ref{tab:table1}). 
Fig.~\ref{fig:correlations_measures} also shows the correlation between our method and single-shell approach in the case of QMSD and QMFD measures. The correlograms for QMSD/QMFD measures clearly show the outliers generated by the single-shell technique, which are not present in the proposed one.

\section{Conclusions}
In this paper, we proposed a new approach based on stretched-exponentials to quantify EAP features such as RTOP, QMSD, and QMFD measures. From the results, it seems that the proposed method reduces the amount of data to be acquired, and therefore it can be clinically feasible. We have to mention that much more thorough validation and comparison are required to bring convincing evidence that comparable results can be obtained with fewer data in our method. In our experiments, we evaluated the method's stability, but the accuracy remains to be investigated in future work. Further, the convergence of the series expansion has to be verified as well. Finally, notice that in situations where the signal cannot be accurately described by the stretched-exponential \cite{novikov2019quantifying}, the scalar measures will be biased. Besides, the proposed method can be generalized to other Q-space factors such as return-to-the-axis or return-to-the-plane probability.


\newpage
%
\bibliographystyle{splncs04}
%
\bibliography{references}
\end{document}